\documentclass[a4paper,pra,twocolumn,superscriptaddress]{revtex4}  

\usepackage[english]{babel}
\usepackage{moreverb}
\usepackage{fancyhdr}
\usepackage{graphicx}
\usepackage{SIunits}
\usepackage{amsmath}
\usepackage{epsfig}
\usepackage[small]{subfigure}

\newcommand{\ks}[0]{\kappa^2}

\newcommand{\kt}[0]{\tilde{\kappa}}

\newcommand{\ha}{\hat{a}}
\newcommand{\hA}{\hat{A}}

\newcommand{\ktot}{\kappa^2_{{\rm tot}}}

\begin{document}
\title{Light Qubit Storage and Retrieval using Macroscopic Atomic
  Ensembles}

\author{J. Sherson } \email{sherson@phys.au.dk}
\affiliation{QUANTOP, Danish
  National Research Foundation Center for Quantum Optics}
\affiliation{Department of Physics
  and Astronomy, University of Aarhus, DK 8000 Aarhus C, Denmark}
\affiliation{Niels Bohr
  Institute, Copenhagen University, Blegdamsvej 17, 2100 Copenhagen
  \O, Denmark}

\author{A. S. S\o rensen } \affiliation{QUANTOP, Danish National Research
  Foundation Center for Quantum Optics} \affiliation{Niels Bohr
  Institute, Copenhagen University, Blegdamsvej 17, 2100 Copenhagen
  \O, Denmark}

\author{J. Fiur\'{a}\v{s}ek}
\affiliation{Department of Optics, Palack\'{y} University,
17. listopadu 50, 77200 Olomouc, Czech Republic}

\author{K. M\o lmer }
\affiliation{QUANTOP, Danish
  National Research Foundation Center for Quantum Optics}
\affiliation{Department of Physics
  and Astronomy, University of Aarhus, DK 8000 Aarhus C, Denmark}

\author{E. S. Polzik }\affiliation{QUANTOP, Danish
  National Research Foundation Center for Quantum Optics}
\affiliation{Niels Bohr
  Institute, Copenhagen University, Blegdamsvej 17, 2100 Copenhagen
  \O, Denmark}

\begin{abstract}
  
  We present an experimentally feasible protocol for the complete
  storage and retrieval of arbitrary light states in an atomic quantum
  memory using the well-established Faraday interaction between light
  and matter. Our protocol relies on multiple passages of a single
  light pulse through the atomic ensemble without the impractical
  requirement of kilometer long delay lines between the passages.
  Furthermore, we introduce a time dependent interaction strength
  which enables storage and retrieval of states with arbitrary pulse
  shapes. The fidelity approaches unity exponentially without squeezed
  or entangled initial states, as illustrated by explicit calculations
  for a photonic qubit.

\end{abstract}


\maketitle   

\begin{figure}[h]
\includegraphics[width=0.40\textwidth]{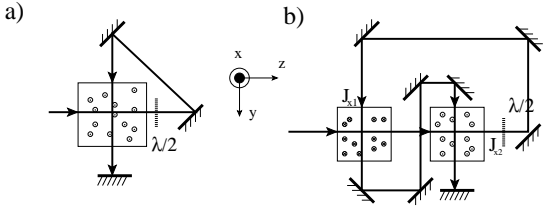}
\caption{\small
  a) Schematic of the multi-pass protocol. The light is sent through
  the atomic sample along the z- and then along the y-direction. After
  the second passage, we may reflect the light back onto the sample to
  improve the performance by two further passages of the light beam.
  b) Schematic of an implementation if the atomic system consists of
  two atomic samples with oppositely oriented mean spin in a
  homogeneous magnetic field as in \cite{mynature04}.
}
 \label{fig:setup}
\end{figure}

Faithful storage and retrieval of an unknown quantum state of
traveling light pulses plays an important role in quantum information
protocols, such as long distance communication
\cite{longdistQcomm,briegel} and quantum computation \cite{klmQcomp}.
Several theoretical proposals for deterministic quantum memory for
light have been put forward using single atoms coupled to high finesse
cavities \cite{CiracNetwork}, as well as atomic ensembles in free
space coupled to light \cite{Kuzmich,contvarbook,LukinQM}.  Recently
\cite{mynature04}, the first quantum storage of weak coherent light
pulses with a higher fidelity than achievable classically was
demonstrated using the off-resonant Faraday rotation of light passing
through a macroscopic atomic ensemble. The subsequent experimental
challenge is to demonstrate retrieval of the quantum state back to a
light pulse following for example the proposal \cite{contvarbook}.
This protocol shows better performance, i.e., higher fidelities, than
achievable classically, but in order to become truly applicable for
future quantum information protocols, fidelities close to unity must
be achieved. This is possible with the protocol, but only if one is
able to prepare highly squeezed initial states, i.e., spin squeezed
atomic states and squeezed input light beams for mapping and retrieval
respectively. The difficulty in producing such squeezed states limits
the realistically achievable fidelity of the quantum memory. In this
Letter we present a new protocol which can reach unity fidelity
without any squeezing. The requirement for this protocol to reach unit
fidelity is sufficient interaction strength, which can be met by a
sufficient number of atoms and photons taking part in the interaction.
Unlike previous proposals our theoretical analysis considers a long
light pulse which is reflected and redirected so that it is on its way
through the sample along different directions at the same time, see
Fig. \ref{fig:setup}.  This dramatically changes the dynamics of the
whole system, and the evolution is no longer governed by the Quantum
Non Demolition (QND) coupling used in the direct mapping protocol
\cite{mynature04} and considered in various atom-light interface
protocols discussed in Refs. \cite{contvarbook,Fiurasek03, Klemens}.
These protocols have assumed sequential passage of the light through
the sample which for pulses of millisecond duration as required in
\cite{mynature04} means that delay lines of hundreds of kilometers
would be necessary.  In addition to presenting a protocol which works
for long pulses, we demonstrate that it is possible to tailor the
interaction strength such that the quantum memory may store and
retrieve incoming light pulse with arbitrary pulse shapes.  Finally,
we note that, although being an effectively continuous variable system
a macroscopic atomic ensemble can be used to store a single qubit and
we shall show explicitly that our protocol enables high fidelity
storage and retrieval of qubit states in the form of superposition
states of 0 and 1 photons.

We investigate a setup with a macroscopic atomic spin angular momentum
oriented along the x-direction \cite{MabuchiSpinSqueezing} or,
equivalently, two samples with opposite macroscopic spin placed in a
constant magnetic field \cite{mynature04}. See Fig. \ref{fig:setup}.
As described in \cite{mynature04,distent} the total angular momentum
along the x-direction, $J_x$, is effectively a classical parameter and
the two remaining components can be rescaled to form operators,
$x_A=J_y/\sqrt{J_x}$ and $p_A=J_z/\sqrt{J_x}$ with canonical
commutation relation $[x_A,p_A]=i$. The light state is composed of a
strong coherent component linearly polarized in the x-direction and
the weak quantum mechanical signal to be stored in the y-polarization.
For this system the Stokes vector component $S_x$ is a macroscopic
classical quantity and we can define the operators
$x_L=S_y/\sqrt{S_x}$ and $p_L=S_z/\sqrt{S_x}$ with commutation
relation $[x_L,p_L]=i$.

As shown in Fig. \ref{fig:setup} our protocol begins with a light
passage along the z-direction.  The off-resonant interaction
Hamiltonian for this process is given by $H\propto p_Ap_L$
\cite{mynature04}.  After a 90 deg. rotation in $x_L$-$p_L$ space the
light is sent through the atomic sample along the y-direction,
creating a $H\propto x_Ax_L$ interaction. To describe the effect of
this interaction we first review the case where the two passes of the
beam occur one after the other \cite{contvarbook}. The output
operators are then given by $p_L^{out}=(1-\kappa^2)p_L^{in}-\kappa
x_A^{in}~,~~~ x_L^{out}=x_L^{in} + \kappa p_A^{in}~,~~~
p_A^{out}=(1-\kappa^2)p_A^{in}-\kappa x_L^{in}~,~~~ x_A^{out}=x_A^{in}
+ \kappa p_L^{in}$, where $\kappa$ is the integrated interaction
strength. $\ks$ is proportional to the total number of atoms and the
total number of photons in the pulse \cite{mygaussent,distent}.
The protocol both maps the light properties on the atoms and the
atomic properties on the outgoing light pulse. For $\kappa=1$, mean
values are faithfully stored whereas uncanceled contributions from
$x_A^{in}$ or $x_L^{in}$ will limit the fidelity of storage or
retrieval of e.g. a coherent state to 88\% \cite{contvarbook}.  As
discussed in the introduction this deficiency can be remedied by
initially squeezing these variables (e.g. $x_A\rightarrow
x_A/\sqrt{\epsilon}$,
$p_A\rightarrow p_A\sqrt{\epsilon}$), in which case the 
fidelity may be increased towards 100\%. This however, represents an
unachievable limit due to the insurmountable difficulty in creating
arbitrarily squeezed states.  Furthermore, as mentioned above, the
requirement of hundreds of kilometer long optical pulses cannot be
avoided for the experimental realization of Ref.  \cite{mynature04}.

%
%
To circumvent this, we propose simultaneous passage of the light
pulses through the atomic medium in two perpendicular directions. We
treat this case by splitting the incoming light pulse into segments of
duration $\tau$ as discussed in \cite{Klemens,mygaussent}.  The
interaction can then be treated sequentially using the coarse grained
Hamiltonian $H_{\tau, i}=\kappa_\tau {p}_{A}{p}_{L,i}$ and $H_{\tau,
  i}^\prime=\kappa_\tau {x}_{A}{x}_{L,i}$.  The crossing of beams in
the atomic sample gives rise to standing wave interference patterns.
Since these will either be stationary or move with a speed of
$c/\sqrt{2}$ in the 45 degree directions with respect to the beams,
atoms with a thermal velocity distribution will effectively average
over these interferences and only experience the traveling wave
components.
By taking the $\tau\to 0$ limit, differential equations are obtained,
which can be solved under the physically reasonable assumption that
the atomic state evolves slowly compared to the time it takes for
light to travel between the two passages \cite{qubitlong}:
\begin{equation}
  \label{eq:atomtwopass}
\begin{aligned}
 x_A (t) &= x_A (0)+ \int_0^t dt' \kt P^{in}_L(t')\\
 p_A (t) &= p_A (0) {\rm e}^{-\int_0^t du \kt^2}
  -  \int_0^t dt'\kt
 {\rm e}^{-\int_{t'}^t du \kt^2} X^{in}_L(t')\\
     X^{out}_L(t)&=X^{in}_L(t)+\kt(t) p_A(t) \\
     P^{out}_L(t)&=P^{in}_L(t)-\kt(t)x_A(t),
\end{aligned}
\end{equation}
where for future use we have defined a time dependent interaction rate
$\kt(t)$ which is characterized by $\kappa^2_{tot} = \int_0^\infty
\kt^2(t)dt$ (time dependence of $\kt$ is implicit in the integrals).
$X_L(t)$ and $P_L(t)$ are the position and momentum operators for the
light segment arriving at time $t$ with commutation relation
$[X_L(t),P_L(t')]=i\delta(t-t')$. As can be seen the solution turns
out to be very similar to the simple QND case reviewed earlier.  In
particular we recognize an asymmetry in which only the initial quantum
fluctuations of one of the atomic quadratures are damped, which is
exactly the reason why squeezing of the initial states is required.

The result in itself is noteworthy, because it gives the first
indication that retrieval of stored states is experimentally feasible
for the system of Ref.  \cite{mynature04}. Indeed we have devised a
protocol, which uses this two pass retrieval in combination with the
single pass storage demonstrated in \cite{mynature04} to achieve an
ideal quantum memory if infinitely squeezed states are available
\cite{qubitlong}. Our main purpose here is to show that one may devise
a protocol which does not require squeezing, and we shall not go into
more detail with this direct mapping\cite{mynature04}/two-pass
retrieval ("1+2") protocol here, except for a comparison of its
performance with the main novel protocol of this paper in which the
light is reflected back through two further passages of the atomic
medium, see Fig.~\ref{fig:setup}.  In the resulting four pass protocol
the light will experience an interaction sequence $pp,~xx,~xx$, and
$pp$, which effectively restores the symmetry between the atomic
quadratures. As in the two pass case a differential equation for the
time evolution of the atomic and light variables can be formulated and
solved to obtain \cite{qubitlong}:
\begin{equation}
  \label{eq:slowvarfourpass}
  \begin{aligned}
 x_A(t) &=   x_A(0){\rm e}^{-2\int_0^t du
 \kt^2}+2\int_0^t dt' \kt {\rm e}^{-2\int_{t'}^t du
 \kt^2} P_L^{in}(t')\\
 p_A(t) &=  p_A(0){\rm e}^{-2\int_0^t du
 \kt^2}-2\int_0^t dt' \kt {\rm e}^{-2\int_{t'}^t du
 \kt^2} X_L^{in}(t')\\
     X^{out}_L(t)&=X^{in}_L(t)+2\kt(t) p_A(t) \\
     P^{out}_L(t)&=P^{in}_L(t)-2\kt(t)x_A(t).
  \end{aligned}
\end{equation}
By introducing the annihilation operators $\ha = (x+ip)/\sqrt{2}$ and
$\hat{A}=(X+iP)/\sqrt{2}$ the entire interaction can be written as a
beam splitter relation:
\begin{equation}
  \label{eq:afourpass}
 \ha_A(t) =   \ha_A(0){\rm e}^{-2\int_0^t du
 \kt^2}-2i\int_0^t dt' \kt {\rm e}^{-2\int_{t'}^t du
 \kt^2} \hA(t').
\end{equation}

\begin{figure}[t]
\centering
  \begin{minipage}[c]{.25\textwidth}
\centering
 \includegraphics[width=\textwidth]{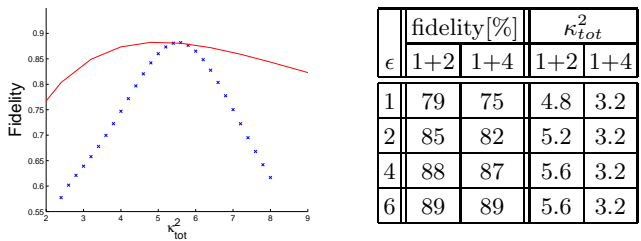}
 \end{minipage}%
  \begin{minipage}[c]{.25\textwidth}
\centering
\begin{tabular}[c]{|c||c|c||c|c|}
\hline
&\multicolumn{2}{l||}{fidelity[\%]}&\multicolumn{2}{c|}{$\kappa_{tot}^2$}\\
\cline{2-5}
$\epsilon$&1+2&1+4&1+2&1+4\\
\hline\hline
1&79&75&4.8&3.2\\
\hline
2&85&82&5.2&3.2\\
\hline
4&88&87&5.6&3.2\\
\hline
6&89&89&5.6&3.2\\
\hline
\end{tabular}
\end{minipage}
 \caption{\small
   Combined storage and retrieval fidelity of a qubit  with
   constant interaction strength for one pass storage and
   two pass retrieval, $\epsilon=4$, (crosses) and four pass storage
   and four pass retrieval (full curve).
   In the table complete storage and retrieval fidelities (obtained
   from the optimum of curves as the ones shown) for two different
   retrieval protocols with varying degrees of squeezing are presented
   along with the optimum values of the accumulated interaction
   strength, $\ks_{tot}$.}
  \label{fig:12_44_fidelvskappatot}
\end{figure}

To quantify the quantum memory performance, we calculate the combined
storage and retrieval fidelity $F=\langle\Psi_{{\rm ideal}}|\rho|
\Psi_{{\rm ideal}}\rangle$, where $\Psi_{{\rm ideal}}$ is the ideal
state and $\rho$ is the retrieved density matrix. We consider the
storage and retrieval of an unknown qubit stored in the zero and one
photon subspace of the light field, and calculate the fidelity
averaged over all orientations on the Bloch sphere. For the "4+4"
protocol, this can be calculated directly from Eq.
(\ref{eq:afourpass}), by noting that imperfect operation results in
admixture of vacuum noise into the mode we are interested in, and the
effect (or an actual source of loss or decoherence) is therefore
fairly easily determined, e.g., by noting that within the 0 and 1
photon subspace (\ref{eq:afourpass}) corresponds to the relation for a
decaying two-level system. The calculations for protocols involving
either a single or a two-pass component are significantly more
complicated involving orthogonal mode decomposition and appropriate
differentiations of the Husimi Q-function and we shall not go into
details here \cite{qubitlong}.  In Fig.
\ref{fig:12_44_fidelvskappatot} we show two examples of the behavior
of the fidelity as a function of the total interaction strength
$\kappa^2_{\rm tot}$ for constant $\kt(t)$. Experimentally $\kt$ is
proportional to the amplitude of the classical light field, and hence
the total number of photons is given by the time integral of $\kt^2$.
As discussed below the noise introduced by spontaneous emission each
time the light passes through the sample is proportional to the number
of photons. Hence, in order to be able to compare protocols involving
different number of passes we compare their performance as a function
of total numbers of photon passes which, e.g, for four pass storage
followed by four pass retrieval is proportional to $\kappa_{{\rm
    tot}}^2=8\int dt \kt^2(t)$. The crosses show the result of the
direct mapping protocol with $\kappa^2_{tot}=2$ and retrieval of light
with a two pass interaction ("1+2") \cite{qubitlong}.  Both the
initial atomic state and the light of the retrieval pulse are
initially squeezed by a factor $\epsilon=4$. The solid curve shows the
fidelity for the "4+4" protocol. The table summarizes the optimal
combined storage and retrieval fidelities at different squeezing
factors for the "1+2" and "1+4" protocols. In each case the total
accumulated interaction strength is also given. We note that with a
moderate amount of squeezing, fidelities well above the classical
boundary of 2/3 can be achieved.  We also note that the "4+4" protocol
does not require any squeezing, and with a moderate $\ks_{\rm
  tot}=4.8$ we get $F=88\%$ with constant coupling strength.
%
%

As a departure from all previous proposals we propose to increase the
achievable degree of fidelity and at the same time facilitate the
storage of light pulses with arbitrary temporal profile by
appropriately varying the interaction constant throughout the pulse.
This is readily achieved since the time dependent interaction strength
$\kt(t)$ is proportional to the amplitude of the strong x-polarized
driving field, which can be varied experimentally using e.g. an
electro-optic modulator (EOM). In order to store an incoming light
pulse with a temporal mode function $f(t)$ so that it is described by
the annihilation operator $ \ha_L^{in}=\int_0^T dt f(t)
\hat{A}^{in}_L(t)$, we want $\ha_A(T)=\ha_L^{in}$ which is equivalent
to
\begin{equation}
  \label{eq:difkappa}
  \frac{1}{\kt}\frac{d\kt}{dt}+2\kt^2=\frac{1}{f}\frac{df}{dt}.
\end{equation}
If we are interested in storing the state of a light pulse described
by any (real) $f(t)$ this differential equation can be solved to
obtain the appropriate shape of the interaction strength $\kt$.  Hence
by suitable tailoring of $\kt$ the protocol allows for ideal storage
(and retrieval) of an incoming light state of arbitrary shape without
any initial squeezing. Note, that the shape of $\kt$ for storage and
retrieval may be different so that the retrieved pulse may be
transformed into any desirable shape.  As a specific example let us
consider a constant field mode of duration $T$ with $f(t)=1/\sqrt{T}$.
In the case of mapping, the initial atomic state is damped
exponentially but so is the early part of the input light pulse.  This
can be counteracted by an increased interaction strength for the front
part of the pulse, and Eq.  (\ref{eq:difkappa}) gives $\kt(t) =
\frac{1}{2\sqrt{t}}$ for the optimal mapping interaction.  When
retrieving a stored state the rear end of the light pulse reads out a
damped atomic state and the divergence needs to be placed at this end
of the pulse.

Another interesting result can be obtained if we let
$\kt(t)=\frac{1}{2\sqrt{t+T}}$, where $T$ is the 
pulse length.  In this case we get a 50/50 beam splitter between the
light and atoms.  This adds to the growing toolbox of interesting
interactions between atomic ensembles and light pulses, which pave the
way for a number of hybrid light-atom protocols, e.g., for
teleportation and entanglement swapping.

\begin{figure}[t]
\includegraphics[width=0.3\textwidth]{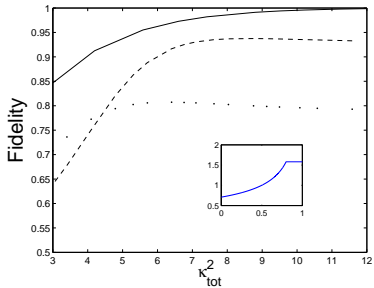}
\caption{\small
  The fidelity of mapping and subsequent retrieval of a qubit with
  different truncations of the interaction strength
  (\ref{eq:difkappa}).  Curves are shown for the one pass direct
  mapping followed by two-pass retrieval with and without squeezing,
  $\epsilon=4$, of the initial states (dashed and dotted respectively)
  and for two truncated four pass pulses (full curve).  Inset: example
  of a truncated interaction strength for retrieval.}
\label{fig:truncated}
\end{figure}

%
%
Our protocol for optimum storage and retrieval of photonic states
requires divergences in the coupling strength, which cannot be
achieved experimentally. To quantify the effect of finite $\kt$, we
assume that we truncate $\kt$ by making it constant close to the
divergences, $1/\sqrt{t}$ and $1/\sqrt{T-t}$ for the mapping and
retrieval respectively, e.g., for the retrieval $\kt =
\min[1/2\sqrt{T-t},\phi]$ where $\phi$ is a chosen constant, see
inset of Fig. \ref{fig:truncated}.  The fidelity for the combined
storage and retrieval can be calculated analytically\cite{qubitlong},
and for large interaction the fidelity approaches unity rapidly:
\begin{equation}
  \label{eq:Ffour}
  F\to 1- \frac{8\sqrt{{\rm e}}-4{\rm e}}{3}\exp(-\kappa^2_{{\rm tot}}/2)~,\qquad \kappa_{\rm tot}\gg1~.
\end{equation}

In Fig. \ref{fig:truncated} we compare the "1+2" and the "4+4"
protocols, both with time-dependent $\kt$.  For the "1+2" we include
curves for which both $x_A^{in}$ and $\tilde{x}_L^{in}$ are squeezed
with a factor $\epsilon=1,4$ . We see that fidelities of the order of
93 \% can be achieved with experimentally feasible squeezing.  As
expected the four pass protocol without squeezing approaches unit
fidelity already for experimentally achievable values of
$\kappa^2_{\rm tot}$, whereas significant squeezing is necessary for
the "1+2" protocol. We thus see that the "4+4" protocol solves the
challenge of feasible high fidelity quantum memory. Note, that this
protocol involves a significant amount of passages so if the atomic
sample is contained within glass cells as in \cite{mynature04} light
reflection loss at each interface would decrease the fidelity. This
decrease, however, can be reduced by adding anti reflection coating to
the glass cells or by having the additional mirrors within the vacuum
chamber in a trapped atom setup.

The effect of spontaneous emission of light by the atoms should also
be addressed.  For a single pass through a gas with resonant optical
density $\alpha$ the spontaneous emission probability per atom $\eta$
is given by $\eta=\ktot/\alpha$\cite{mygaussent}. For several passes
the probability of spontaneous emission is proportional to the number
of passes, which is why we multiply the single passage $\ktot$ by the
number of passes when comparing different schemes. Because of the
spontaneous emission it is important to keep $\ktot$ fairly small.
The effect of spontaneous emission will reduce
the fidelity of the protocol by an amount proportional to
$\eta$, so that in combination with Eq. (\ref{eq:Ffour}) the total
error $(1-F$) is of the order of $\sim \exp(-\alpha\eta)+\eta$.
Optimizing this expression with respect to $\eta$ we find that the
error scales as $\log(\alpha)/\alpha$. It is thus advantageous to make
$\alpha$ as large as possible, something which is readily achieved in
trapped cold samples \cite{MabuchiSpinSqueezing}. Here large optical
depths are achieved for ultra compact samples of atoms, thus preserving
the prospect of future miniaturization.

The present protocols can also be extended to employ two atomic memory
units (ensembles) $A$ and $B$ to store or retrieve a qubit originally
represented by a single photon in two modes $L$ and $M$,
$|\psi\rangle_{LM}=\alpha|10\rangle_{LM}+\beta|01\rangle_{LM}$.  This
is important since this encoding is used in a vast majority of current
experiments on optical quantum information processing with discrete
variables.  The fidelity of storage and/or retrieval can be calculated
using similar techniques as for the single-mode qubit
\cite{qubitlong}.
%
%
%
%
%

In conclusion, existing protocols for using the off-resonant Faraday
interaction to retrieve an unknown light state stored in a macroscopic
atomic sample \cite{mynature04} are very impractical because they
assume sequential passage of the light beam which requires
prohibitively long delay lines. Furthermore, even if the protocols
could be realized experimentally they predict unity fidelity only in
the limit of infinitely squeezed input states.

Our solution to both of these problems involves three novel features.
First of all, we have solved the dynamics arising when the probe pulse
travels through the atomic sample in two orthogonal directions
simultaneously. This we applied to a novel scheme involving four
passages of the light. Combined with the third new component: a time
varying interaction strength, this protocol can be used for storage or
retrieval symmetrically and works with fidelity exponentially
approaching unity with increased interaction strength.  No squeezing
is required and since the interaction strength in the Faraday
interaction depends on the number of atoms and the number of photons
there is no fundamental limit to the achievable fidelity for this
protocol.  As a final remark, we note that the feasibility of the
protocol is accentuated by the fact that all basic components are in
place already. In experiments like \cite{mynature04} one only has to
add eight mirrors as illustrated in Fig. \ref{fig:setup}.

 \indent This work was supported by the EU under project
COVAQIAL (FP6-511004), by the Danish Natural Science Research Council,
and by the Danish National Research Foundation.  JF acknowledges
support by M\v{S}MT (MSM6198959213) and GA\v{C}R (202/05/0486).

\bibliographystyle{apsrev}
\bibliography{paper}

\begin{thebibliography}{14}
\expandafter\ifx\csname natexlab\endcsname\relax\def\natexlab#1{#1}\fi
\expandafter\ifx\csname bibnamefont\endcsname\relax
  \def\bibnamefont#1{#1}\fi
\expandafter\ifx\csname bibfnamefont\endcsname\relax
  \def\bibfnamefont#1{#1}\fi
\expandafter\ifx\csname citenamefont\endcsname\relax
  \def\citenamefont#1{#1}\fi
\expandafter\ifx\csname url\endcsname\relax
  \def\url#1{\texttt{#1}}\fi
\expandafter\ifx\csname urlprefix\endcsname\relax\def\urlprefix{URL }\fi
\providecommand{\bibinfo}[2]{#2}
\providecommand{\eprint}[2][]{\url{#2}}

\bibitem[{\citenamefont{Duan et~al.}(2001)\citenamefont{Duan, Lukin, Cirac, and
  Zoller}}]{longdistQcomm}
\bibinfo{author}{\bibfnamefont{L.-M.} \bibnamefont{Duan}},
  \bibinfo{author}{\bibfnamefont{M.~D.} \bibnamefont{Lukin}},
  \bibinfo{author}{\bibfnamefont{J.~I.} \bibnamefont{Cirac}}, \bibnamefont{and}
  \bibinfo{author}{\bibfnamefont{P.}~\bibnamefont{Zoller}},
  \bibinfo{journal}{Nature} \textbf{\bibinfo{volume}{414}},
  \bibinfo{pages}{413} (\bibinfo{year}{2001}).

\bibitem[{\citenamefont{Briegel et~al.}(1998)\citenamefont{Briegel, Dür, Cirac,
  and Zoller}}]{briegel}
\bibinfo{author}{\bibfnamefont{H.~J.} \bibnamefont{Briegel}},
  \bibinfo{author}{\bibfnamefont{W.}~\bibnamefont{Dür}},
  \bibinfo{author}{\bibfnamefont{J.~I.} \bibnamefont{Cirac}}, \bibnamefont{and}
  \bibinfo{author}{\bibfnamefont{P.}~\bibnamefont{Zoller}},
  \bibinfo{journal}{Phys. Rev. Lett.} \textbf{\bibinfo{volume}{81}},
  \bibinfo{pages}{5932} (\bibinfo{year}{1998}).

\bibitem[{\citenamefont{Knill et~al.}(2001)\citenamefont{Knill, Laflamme, and
  Milburn}}]{klmQcomp}
\bibinfo{author}{\bibfnamefont{E.}~\bibnamefont{Knill}},
  \bibinfo{author}{\bibfnamefont{R.}~\bibnamefont{Laflamme}}, \bibnamefont{and}
  \bibinfo{author}{\bibfnamefont{G.~J.} \bibnamefont{Milburn}},
  \bibinfo{journal}{Nature} \textbf{\bibinfo{volume}{409}}, \bibinfo{pages}{46}
  (\bibinfo{year}{2001}).

\bibitem[{\citenamefont{Cirac et~al.}(1997)\citenamefont{Cirac, Zoller, Kimble,
  and Mabuchi}}]{CiracNetwork}
\bibinfo{author}{\bibfnamefont{J.~I.} \bibnamefont{Cirac}},
  \bibinfo{author}{\bibfnamefont{P.}~\bibnamefont{Zoller}},
  \bibinfo{author}{\bibfnamefont{H.~J.} \bibnamefont{Kimble}},
  \bibnamefont{and} \bibinfo{author}{\bibfnamefont{H.}~\bibnamefont{Mabuchi}},
  \bibinfo{journal}{Phys. Rev. Lett.} \textbf{\bibinfo{volume}{78}},
  \bibinfo{pages}{3221} (\bibinfo{year}{1997}).

\bibitem[{\citenamefont{Kuzmich and Polzik}(2000)}]{Kuzmich}
\bibinfo{author}{\bibfnamefont{A.}~\bibnamefont{Kuzmich}} \bibnamefont{and}
  \bibinfo{author}{\bibfnamefont{E.~S.} \bibnamefont{Polzik}},
  \bibinfo{journal}{Phys. Rev. Lett.} \textbf{\bibinfo{volume}{85}},
  \bibinfo{pages}{5639} (\bibinfo{year}{2000}).

\bibitem[{\citenamefont{Kuzmich and Polzik}(2003)}]{contvarbook}
\bibinfo{author}{\bibfnamefont{A.}~\bibnamefont{Kuzmich}} \bibnamefont{and}
  \bibinfo{author}{\bibfnamefont{E.~S.} \bibnamefont{Polzik}},
  \emph{\bibinfo{title}{Quantum Infomation with Continuous Variables}}
  (\bibinfo{publisher}{Kluwer}, \bibinfo{address}{Dordrecht},
  \bibinfo{year}{2003}), pp. \bibinfo{pages}{231--265}, \bibinfo{note}{eds. S.
  L. Braunstein and A. K. Pati}.

\bibitem[{\citenamefont{Fleischauer and Lukin}(2002)}]{LukinQM}
\bibinfo{author}{\bibfnamefont{M.}~\bibnamefont{Fleischauer}} \bibnamefont{and}
  \bibinfo{author}{\bibfnamefont{M.~D.} \bibnamefont{Lukin}},
  \bibinfo{journal}{Phys. Rev. A} \textbf{\bibinfo{volume}{65}},
  \bibinfo{pages}{022314} (\bibinfo{year}{2002}).

\bibitem[{\citenamefont{Julsgaard et~al.}(2004)\citenamefont{Julsgaard,
  Sherson, Cirac, Fiur\'{a}\v{s}ek, and Polzik}}]{mynature04}
\bibinfo{author}{\bibfnamefont{B.}~\bibnamefont{Julsgaard}},
  \bibinfo{author}{\bibfnamefont{J.}~\bibnamefont{Sherson}},
  \bibinfo{author}{\bibfnamefont{J.}~\bibnamefont{Cirac}},
  \bibinfo{author}{\bibfnamefont{J.}~\bibnamefont{Fiur\'{a}\v{s}ek}},
  \bibnamefont{and} \bibinfo{author}{\bibfnamefont{E.}~\bibnamefont{Polzik}},
  \bibinfo{journal}{Nature} \textbf{\bibinfo{volume}{432}},
  \bibinfo{pages}{482} (\bibinfo{year}{2004}).

\bibitem[{\citenamefont{Fiur\'{a}\v{s}ek}(2003)}]{Fiurasek03}
\bibinfo{author}{\bibfnamefont{J.}~\bibnamefont{Fiur\'{a}\v{s}ek}},
  \bibinfo{journal}{Phys. Rev. A} \textbf{\bibinfo{volume}{68}},
  \bibinfo{pages}{022304} (\bibinfo{year}{2003}).

\bibitem[{\citenamefont{Hammerer et~al.}(2004)\citenamefont{Hammerer, Mølmer,
  Polzik, and Cirac}}]{Klemens}
\bibinfo{author}{\bibfnamefont{K.}~\bibnamefont{Hammerer}},
  \bibinfo{author}{\bibfnamefont{K.}~\bibnamefont{Mølmer}},
  \bibinfo{author}{\bibfnamefont{E.~S.} \bibnamefont{Polzik}},
  \bibnamefont{and} \bibinfo{author}{\bibfnamefont{J.~I.} \bibnamefont{Cirac}},
  \bibinfo{journal}{Phys. Rev. A} \textbf{\bibinfo{volume}{70}},
  \bibinfo{pages}{044304} (\bibinfo{year}{2004}).

\bibitem[{\citenamefont{Geremia et~al.}(2004)\citenamefont{Geremia, Stockton,
  and Mabuchi}}]{MabuchiSpinSqueezing}
\bibinfo{author}{\bibfnamefont{J.}~\bibnamefont{Geremia}},
  \bibinfo{author}{\bibfnamefont{J.}~\bibnamefont{Stockton}}, \bibnamefont{and}
  \bibinfo{author}{\bibfnamefont{H.}~\bibnamefont{Mabuchi}},
  \bibinfo{journal}{Science} \textbf{\bibinfo{volume}{304}},
  \bibinfo{pages}{270} (\bibinfo{year}{2004}).

\bibitem[{\citenamefont{Sherson et~al.}(2004)\citenamefont{Sherson, Julsgaard,
  and Polzik}}]{distent}
\bibinfo{author}{\bibfnamefont{J.}~\bibnamefont{Sherson}},
  \bibinfo{author}{\bibfnamefont{B.}~\bibnamefont{Julsgaard}},
  \bibnamefont{and} \bibinfo{author}{\bibfnamefont{E.~S.} \bibnamefont{Polzik}}
  (\bibinfo{year}{2004}), \bibinfo{note}{quant-ph/0408146}.

\bibitem[{\citenamefont{Sherson and Mølmer}(2005)}]{mygaussent}
\bibinfo{author}{\bibfnamefont{J.}~\bibnamefont{Sherson}} \bibnamefont{and}
  \bibinfo{author}{\bibfnamefont{K.}~\bibnamefont{Mølmer}},
  \bibinfo{journal}{Phys. Rev. A} \textbf{\bibinfo{volume}{71}},
  \bibinfo{pages}{033813} (\bibinfo{year}{2005}).

\bibitem[{qub()}]{qubitlong}
\bibinfo{note}{J. Sherson et al, In preparation}.

\end{thebibliography}
\end{document}